# Intrinsic circularly-polarized exciton emission in a twisted van-der-Waals heterostructure

J. Michl[1], S.A. Tarasenko[2], F. Lohof[3], G. Gies[3], M. von Helversen[4], R. Sailus[5], S. Tongay[5], T. Taniguchi[6], K. Watanabe[7], T. Heindel[4], S. Reitzenstein[4], T. Shubina[2], S. Höfling[1], C. Anton-Solanas[1,8], C. Schneider[8]

[1]*Technische Physik, Physikalische Institut and Wilhelm Conrad Röntgen-Center for Complex Material Systems, Universität Würzburg, Am Hubland, D-97074 Würzburg, Germany*
[2]*Ioffe Institute, 194021 St. Petersburg, Russia*
[3]*Institute for Theoretical Physics, University of Bremen, D-28334 Bremen, Germany*
[4]*Institute of Solid-State Physics, Technische Universität Berlin, D-10623 Berlin, Germany.*
[5]*School for Engineering of Matter, Transport, and Energy, Arizona State University, Tempe, AZ, USA.*
[6]*International Center for Materials Nanoarchitectonics, National Institute for Materials Science, 1-1 Namiki, Tsukuba 305-0044, Japan*
[7]*Research Center for Functional Materials, National Institute for Materials Science, 1-1 Namiki, Tsukuba 305-0044, Japan*
[8]*Institute of Physics, Carl von Ossietzky University, 26129 Oldenburg, Germany.*

**The investigation of excitons in van-der-Waals heterostructures has led to profound insights into the interplay of crystal symmetries and fundamental effects of light-matter coupling. In particular, the polarization selection rules in undistorted, slightly twisted heterostructures of $MoSe_2/WSe_2$ were found to be connected with the Moiré superlattice. Here, we report the emergence of a significant degree of circular polarization of excitons in such a hetero-structure upon non-resonant driving with a linearly polarized laser. The effect is present at zero magnetic field, and sensibly reacts on perpendicularly applied magnetic field. The giant magnitude of polarization, which cannot be explained by conventional birefringence or optical activity of the twisted lattice, suggests a kinematic origin arising from an emergent pyromagnetic symmetry in our structure, which we exploit to gain insight into the microscopic processes of our device.**

## Introduction

Controlling the polarization of the emitted light in layered semiconductor structures is a delicate task in solid-state photonics, being of greatest importance in optoelectronic applications. The most known device to control the polarization of light is the standard wave plate, whose working principle relies on the effect of optical birefringence in anisotropic materials. However, most transparent materials used for such optical or near infrared applications have limited optical anisotropy, rendering wave plates inherently bulky and unsuitable for microscale photonic integration. More recently, it was found that chiral nanostructures [1], microcavities [2] and photonic crystal slabs with chiral symmetry [3,4] cause strong polarization of light from unpolarized emitters. By implementing such chiral photonic structures, it became possible to realize nano- and micro-lasers [5], which intrinsically feature a strong degree of circular polarization without the need for external magnetic fields that is also a key point for chip technology. The circular polarization effect in such chiral structures is based on a special engineering of the electromagnetic field in photonic cavities with reduced symmetry, for example, by forming polarizing pattern, whose elements has a shape of gammadions [6]. However, the lateral dimensions of these patterned structures are still too large for on-chip next-generation photonics.

Recently, much attention has been paid to the study of atomically thin transition metal dichalcogenides (TMDCs), since they can absorb up to 10% of incident light at less than 1-nm thickness [7]. Furthermore, monolayers of TMDCs have unique optical properties, such as the locking of spin and valley degree of freedom, which arises from the broken inversion symmetry of the hexagonal crystals [7]. It is noteworthy that without the application of external magnetic fields, the photoluminescence of 2D TMDCs crystals does not exhibit any degree of optical circular polarization (DCP).

In this work, we investigate whether systems of ultimate thinness, namely an atomically thin chiral van-der-Waals heterostructure composed of two different TMDC layers, can efficiently convert linearly polarized or unpolarized light into light featuring a degree of circular polarization. Our study follows the theoretical prediction by Poshakinskiy et al., which suggested optical activity in chiral stacks of TMDC [8]. Our experiment, which is carried out using a twisted bilayer of $MoSe_2/WSe_2$, reveals compelling features of elliptical polarization generated by Moiré excitons. Surprisingly, we observe DCP orders of magnitude larger than $\sim d/\lambda$, expected from optical activity of such a chiral stack, where $d$ is the bilayer width and $\lambda$ is the light wavelength. Our analysis suggests a kinetic origin of exciton circular polarization. The theoretical model predicts DCP up to 0.5 (50 %) which is in agreement with the maximal experimental DCP of 0.44 ± 0.02, which we detect in our experiments.

**Experimental results**

We utilize the dry stamping method [9] in a home-built microscope to assemble the van-der-Waals heterostructure composed of a nominally 10 nm thin layer of hexagonal boron nitrite (hBN, deposited on a $SiO_2$ substrate, see dashed yellow line depicting the limits of the hBN sheet), a monolayer of $MoSe_2$ and a final $WSe_2$ monolayer. The corresponding green and red areas in Fig. 1(a) outline the monolayers in the optical microscope image of the studied sample. These layers were isolated via mechanical exfoliation from chemical vapor deposition grown crystals. The corresponding vertical stacking order is sketched in Fig. 1(b), following the same color code for the different materials.

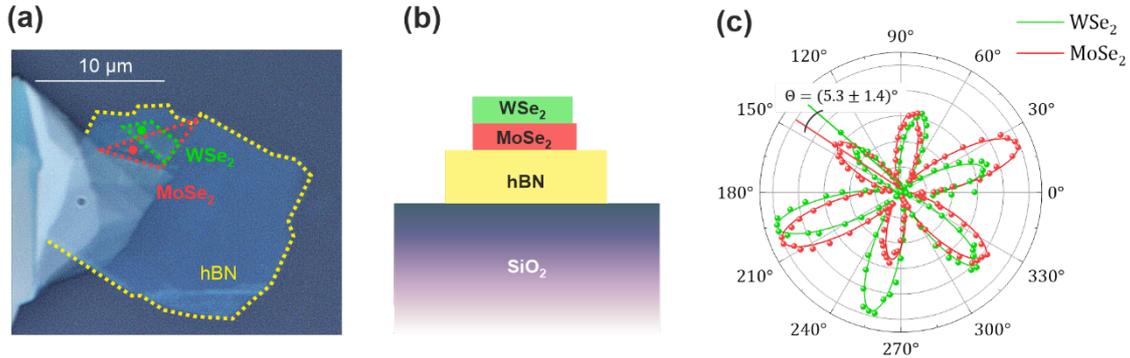

**Fig. 1**. **Description of the sample and determination of the heterostructure twist angle**. (a) Microscope image of the sample, the $hBN/WSe_2/MoSe_2$ flake limits are marked in a dashed yellow/green/red line. (b) Sketch of the corresponding stacking order of the sample. (c) Linear polarization analysis of the second harmonic generated light from freestanding monolayer regions of $WSe_2$ and $MoSe_2$ (see green and red dots in panel (a), respectively, where the measure is performed), the polar plots correspond to the normalized intensities versus the analyzed linear polarization orientation. The relative rotation between the two plots yields a twist angle of (5.3±1.4) deg.

We perform second harmonic generation measurements at room temperature to determine the twist angle between the two $WSe_2/MoSe_2$ sheets [10]. We use a tunable mode locked laser (tuned at 1480 nm) that excites isolated $WSe_2$ and $MoSe_2$ regions (positions indicated in Fig. 1 (a) by a green and a red dot) with a linear polarization oriented at a tunable angle $\varphi$. We detect the second harmonic generated signal at 740 nm (1.676 eV) as a function of the angle $\varphi$ of the incoming laser linear polarization, yielding the polar plots shown in Fig. 1(c). The phase shift between the two plots reveals the twist angle $\theta = (5.3 \pm 1.4)°$ between the $WSe_2$ and $MoSe_2$ layers. Importantly, we note that this angle can locally vary upon the formation of domains with lowered symmetry [11], and we expect this angle to mark an upper boundary in the region where the two monolayers are overlapping.

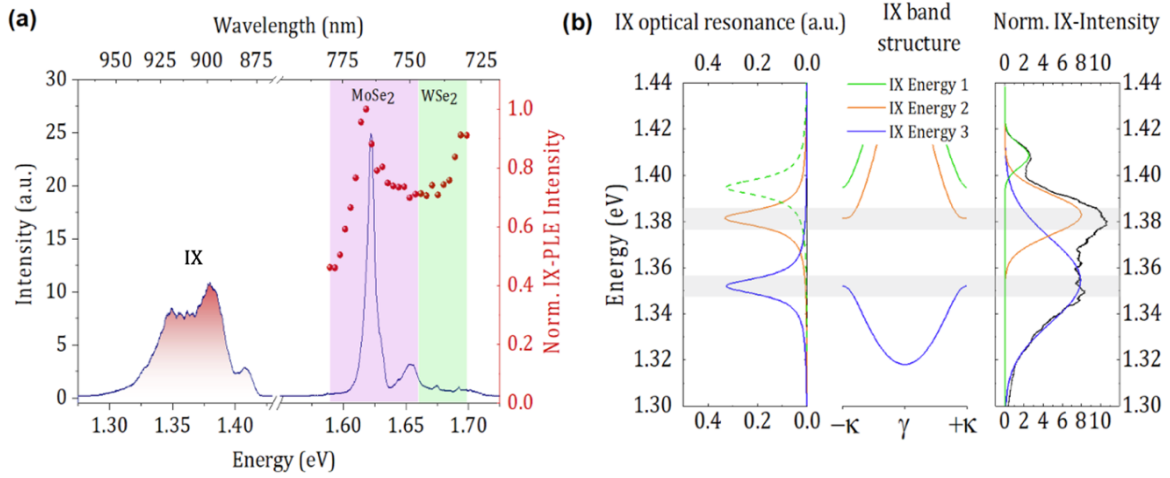

**Fig. 2. Photoluminescence of intra- and inter-layer excitons.** (a) Comparison of photoluminescence of interlayer (see label IX) and intralayer (see labels WSe$_2$- and MoSe$_2$-X) excitons with an excitation power of 0.41 mW and at 1.8 K. The red dots over the intralayer excitons spectrum represent the PLE spectrum of the interlayer exciton, revealing two resonances where carrier transfer occurs. (b) [left] Plot of the optical resonances obtained from the calculated band-structure. [center] Corresponding IX band-structure. [right] Experimental spectrum reproduced from panel (a) presenting with Lorentzian fits, matching the energies of the resonances and dispersion at the K-points of the corresponding reduced Brillouin-zone.

Having established the rotational alignment of the two atomically thin layers, in a next step, we study the intrinsic optical properties of our system at cryogenic temperatures (1.8 K). First, we excite the sample with a 532 nm laser (focused on a spot size of 3 µm diameter), and collect the emitted photoluminescence (PL) over a broad spectral range. In Fig. 2(a), we plot the PL spectrum, which is characterized by two sets of pronounced features arising from the interlayer indirect exciton (IX, energy emission around 1.30-1.40 eV) and the intralayer excitons (energy emission around 1.60-1.70 eV).

The intralayer luminescence from both MoSe$_2$ and WSe$_2$ excitons is composed of neutral and charged excitonic complexes (see Fig. 2(a), where the different excitonic complexes are labelled in their spectrum). In contrast, at lower emission energies we observe a distinct multi-peak signature that we attribute to interlayer excitons formed in the WSe$_2$/MoSe$_2$ heterobilayer. As we explain later in the manuscript, these peaks emerge as a result of trapping excitons in the quasi-periodic Moiré potential [12,13].

In order to confirm the hypothesis of the formation of interlayer excitons by rapid carrier transfer between the two monolayers, we study the PL excitation (PLE) spectrum of the IX emission for excitation energies between 1.59 eV and 1.70 eV. The overall PLE-intensities of the IX are determined by integration of the PLE spectra between 1.27 eV and 1.44 eV and they exhibit two maxima at ~1.62 eV and ~1.69 eV, see red dots in Fig.2(b). The PLE was implemented with a tunable continuous wave laser, using a constant pump power of 5.0 µW across all the scanning excitation energies. The PLE spectrum suggests efficient carrier separation within the heterobilayer and formation of IX complexes following both a photo-injected excitation in the WSe$_2$ layer (hole tunneling) and the MoSe$_2$ layer (electron tunneling). Indeed, slight spectral shifts between the excitonic PL from the monolayers and the PLE resonances in WSe$_2$/MoSe$_2$ stacks have been observed before [13,14] and can be attributed to renormalization phenomena related to dielectric screening and strain in the stacking procedure.

Earlier works have attributed the emergence of a multi-peak structure in the PL signal of slightly rotated van-der-Waals heterostructures to the emergence of Moiré trapped excitons [12]. To support this interpretation in our sample, we model the energy spectrum of IX excitons in the presence of a Moiré potential using a low-energy continuum model [15,16]. In this framework, an effective exciton Hamiltonian is formulated that describes the center-of-mass motion of the interlayer excitons in the twist-angle dependent Moiré potential landscape

$$H = -\frac{\hbar^2}{2M}\Delta_R + V^M(R).$$  [1]

The first term describes the center-of-mass kinetic energy of the IXs with total mass $M = m_e + m_h$, and $V^M(R)$ is the Moiré potential that the IXs experience. More details on the modelling can be found in the supplementary section.

In Fig. 2(b) we plot the IX Moiré exciton band structure as a function of momentum in the Moiré mini-Brillouin zone. At a rotation angle of 5° (close to our experimental case), Eq. (1) yields a set of optical resonances at K/K- that are separated by 20-40 meV [17]. For the identification of the resonances in the emission spectrum, we consider the selection rules on the basis of dipole matrix elements for $\sigma^\pm$ and $z$ polarization, according to which the lower two resonances are bright for normal incidence. We associate these with the two lower-energy peaks in the measured PL spectrum in the right panel of Fig. 2(b). The third Moiré-exciton band obeys selection rules that corresponds to linearly polarized emission and a dipole normal to the bilayer plane. It is intrinsically dark for normal incidence, but contributes z-polarized emission as we can infer from evaluating the optical response for parallel incidence. Acknowledging the fact that the wide aperture of the experimental setup collects some light away from normal incidence, we attribute the emission from the third peak at about 1.41eV to contributions from the third Moiré exciton band.

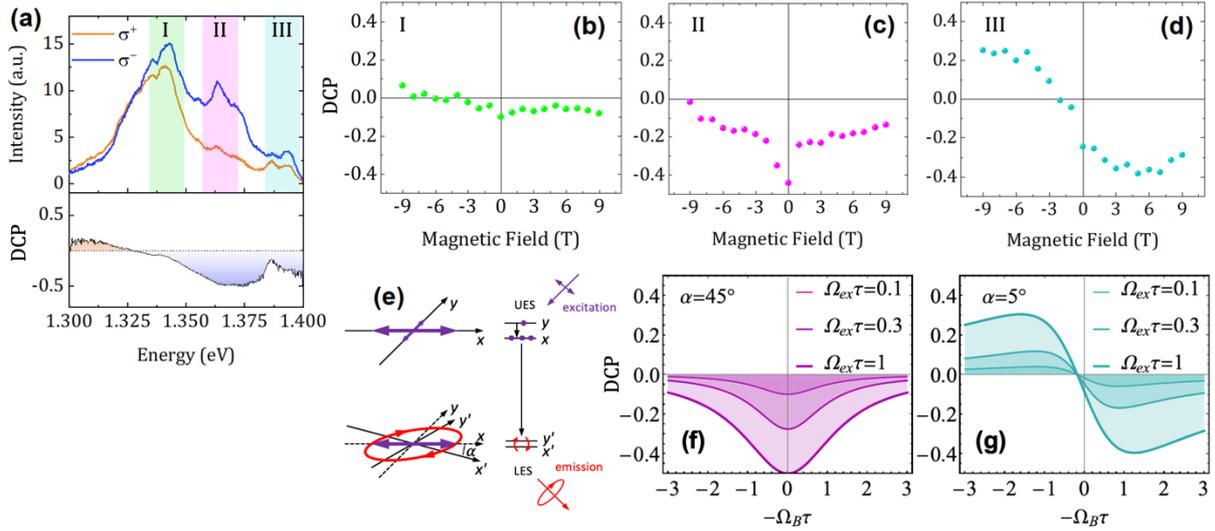

**Fig. 3. Circular polarization analysis of the IX spectrum.** (a) Direct comparison of two PL spectra: The sample is driven by a linear polarized laser, and luminescence is recorded under $\sigma^+$ and $\sigma^-$ circular-polarization detection (orange and blue traces, respectively). The lower panel depicts the corresponding DCP as a function of the emission energy. Three energies regions I, II, and III are marked in the spectrum in green, magenta and cyan colors. (b-d) DCP values versus magnetic field from the three energy regions, respectively. The DCP values are obtained averaging the DCP at each energy in a spectral interval of 20 meV around the marked energy regions (the error bars are the corresponding standard deviation from the average value). (e) Microscopic model of kinetic spin polarization of excitons yielding circular polarized emission. Exciton alignment along $x$ in longitudinal-transverse-split upper exciton state (UES) followed by quantum beats in rotated longitudinal-transverse-split lower exciton state (LES) leads to exciton spin polarization. (f) and (g) Circular polarization of excitons as a function of $\Omega_B\tau$, represented according to Eq. (4) for angles of the longitudinal-transverse eigen-frames of $\alpha = 45$ and 5 degrees and $\Omega_{ex}\tau = 0.1, 0.3, 1$ (see the legend in the corresponding panels).

While the selection rules of the Moiré exciton peaks under circularly-polarized laser driving have been investigated [12,18] we are interested in the intrinsic possibility of IX to emit light carrying a significant degree of circular polarization. Therefore, we excite the system with a CW laser (excitation power ~0.5 µW, and spot size 3 µm diameter) resonant with the $WSe_2$ transition (730 nm) using *linearly* polarized light. We analyze the degree of circular polarization (DCP) of the IX-PL emission.

In Fig. 3 (a), we plot two PL spectra of the Moiré exciton band, detecting $\sigma^+$ and $\sigma^-$ polarized light. The most remarkable feature is captured within the central PL peak, located in region II. It clearly displays a predominance of the $\sigma^-$ polarized component. The observation is visualized by calculating the DCP, calculated as $(I_{\sigma^+} - I_{\sigma^-})/(I_{\sigma^+} + I_{\sigma^-})$, which is displayed in the bottom panel. For energies around 1.363 eV the DCP value reaches a minimum value of -0.44±0.02.

To gain deeper insight in the polarization behavior of our sample, we acquire PL spectra at different magnetic fields. For each magnetic field value, the PL emission is measured in $\sigma^+$ and $\sigma^-$ detection, and subsequently the DCP is obtained. For the sake of clarity, we subdivide the signal in three spectral regions I,II,III, corresponding to the three assigned Moiré exciton peaks (see labels in Fig. 3(a) marked with colored rectangles). At these energies, the DCP is extracted as a function of the applied magnetic field. The energy region I has a small intrinsic DCP for all tested magnetic fields, see Fig. 3(b). We observe that that the DCP attributed to the energy region III can be tuned by the magnetic field to change the DCP sign, see Fig. 3(d). In stark contrast, the central peak in region II, Fig. 3(c), behaves fundamentally different. The absolute DCP value is maximal for *B* = 0 and decreases independent of the sign of the applied magnetic fields.

**Discussion**

The giant circular polarization of exciton emission, which is certainly unexpected from non-magnetic samples and significantly exceeds the expected polarization resulting from a natural birefringence and optical activity, and its unusual dependence on magnetic field suggest a qualitatively different mechanism of exciton spin polarization.

A possibly way to explain our results arises from a kinetic spin polarization of excitons in twisted TMDC heterobilayers as discussed in the following: In analogy to the experimental conditions, our model considers that excitons are optically injected into an upper exciton state (UES), relax into lower exciton state (LES), and then recombine emitting photons, as illustrated in Fig. 3(e). Like in our experimental conditions, the upper state is represented by the resonantly driven WSe$_2$ intralayer transition, which then relaxes to the IX state, which is the LES.

The upper exciton state consists of two sub-levels corresponding to the excitons linearly polarized along some orthogonal $x$ and $y$ axes (cf. the sketch depicted in Fig. 3(e)). We assume that these sub-levels are longitudinal-transverse split due to the anisotropic exchange interaction between electrons and holes. The anisotropic exchange interaction can originate from local in-plane strain or structural imperfections and is known to be large in TMCD layers [19–23]. In the case of large splitting compared to the thermal energy and fast spin relaxation compared to the lifetime in the upper exciton state, the excitons polarize along the $x$ axis.

At the second step, linearly polarized excitons relax to the final LES from which the light emission occurs, i.e. the IX state. The LES is also composed of a doublet consisting of two orthogonal linearly polarized exciton states. Importantly, in twisted bilayers, the eigen-axes of the LES duplet $x'$ and $y'$ do not coincide with the eigen-axes of the UES doublet. As a result, the excitons experience quantum beats in the lower-energy states inducing a circular polarization.

In the pseudospin representation, the average exciton polarization in the lower-energy state is described in the Poincare sphere by the vector $s = S/N$, where $S$ is the vector of the total exciton polarization and $N$ is the number of excitons. The components $(s_x, s_y, s_z) = (DLP, DDP, DCP)$ describe the Stokes polarization parameters of excitons: the linear polarization in the $(x, y)$ plane, the linear polarization in the diagonals of the $x$ and $y$ axes, and the circular polarization, respectively. The vector $S$ satisfies the equation

$$\frac{d\mathbf{S}}{dt} = \mathbf{\Omega} \times \mathbf{S} + \mathbf{G} - \frac{\mathbf{S}}{\tau}, \qquad [2]$$

where $\mathbf{\Omega} = (\Omega_{ex} \cos 2\alpha, \Omega_{ex} \sin 2\alpha, \Omega_B)$ is the effective Lamor frequency corresponding to the anisotriopic exchange interaction in the LES, $\Omega_{ex} = \Delta_{ex}/\hbar$, $\Delta_{ex}$ is the exchange splitting, $\alpha$ is the angle between the $x$ and $x'$ axes (i.e. the rotation angle between the eigen-frames, see Fig 3(e), $\Omega_B = g\mu_B B/\hbar$ is the frequency of spin precession in the external magnetic field $\mathbf{B} \parallel z$, $g$ is the effective exciton $g$–factor, $\mathbf{G}$ is the exciton polarization along $x$ in the intermediate state, $1/\tau = 1/\tau_0 + 1/\tau_s$, with $\tau_0$ and $\tau_s$ the recombination time and spin relaxation time of excitons in the final state, respectively. We do not consider thermalization of excitons in the lower state, since it does not lead to build-up of circular polarization at zero magnetic field.

The number of excitons $N$ is found from the equation $dN/dt = g - N/\tau_0$, where $g$ is the exciton incoming from the upper to the lower state. The steady-state solution of Eq. (2) has the form, Eq. [3]:

$$\mathbf{S} = \frac{\tau \mathbf{G}^2 + \tau^2 \mathbf{\Omega} \times \mathbf{G} + \tau^3 \mathbf{\Omega}(\mathbf{\Omega} \cdot \mathbf{G})}{1 + \Omega^2 \tau^2}, \qquad [3]$$

yielding the Stokes parameters (degree of linear, diagonal and circular polarization, DLP, DDP and DCP, respectively) of the excitonic emission

$$\text{DLP} = \frac{1 + (\Omega_{ex}^2 \tau^2/2)(1 + \cos 4\alpha)}{1 + (\Omega_{ex}^2 + \Omega_B^2)\tau^2} \frac{\tau}{\tau_0} P_0$$

$$\text{DDP} = \frac{(\Omega_{ex}^2 \tau^2/2) \sin 4\alpha + \Omega_B \tau}{1 + (\Omega_{ex}^2 + \Omega_B^2)\tau^2} \frac{\tau}{\tau_0} P_0$$

$$\text{DCP} = \frac{-\Omega_{ex} \tau \sin 2\alpha + \Omega_{ex} \Omega_B \tau^2 \cos 2\alpha}{1 + (\Omega_{ex}^2 + \Omega_B^2)\tau^2} \frac{\tau}{\tau_0} P_0, \qquad [4]$$

where $P_0 = G/g$ is the average linear polarization of excitons in the upper state. The DCP of excitons in the final state can be notable: In the optimal case of $P_0 = 1$, $\alpha = \pi/4$, $\Omega_{ex}\tau = 1$, and $\tau_0 \ll \tau_s$, as plotted in Fig. 3(f), the DCP reaches 50%. The external out-of-plane magnetic field $\mathbf{B}$ affects the kinetic spin polarization, see Eqs. (3-5). In the structures with $\alpha = \pi/4$, the DCP has a maximum at zero magnetic field and is suppressed by the field $\mathbf{B}$. This is furthermore evident in the representation in Fig. 3(g), which reproduces the fundamental behavior of the captured polarization of our Moiré excitons in the energy range 1.35 – 1.38 eV shown in Fig. 3(c).

In non-chiral structures, where $\alpha = 0$ or $\pi/2$, the DCP vanishes at zero magnetic field, rises with field at small fields, reaches a maximum at $\Omega_B \tau = 1$, and then decays, following the standard Lorentz curve. The field dependence of the spin polarization is thus controlled by the parameter $\Omega_B \tau$, and the spin polarization can be much higher than the thermal spin polarization controlled by the ratio between the Zeeman splitting and the thermal energy.

For intermediate conditions at slightly rotated eigen-frames (e.g. 5°), we reach the scenario which is plotted in Fig. 3(g), and which succeeds to reproduce the polarization response of the high energy peak in our spectrum. Here, we observe a modest degree of circular polarization at zero magnetic field, which asymmetrically increases or inverts at negative/positive applied fields.

We note that the emergence of exciton spin polarization $S_z$ (in the absence of an external magnetic field $\mathbf{B}$ and not related to the polarization of optical pump) is allowed by the $C_3$ point-group symmetry of heterobilayers. In this pyromagnetic group, the $z$ component of an axial vector is invariant and, therefore, $S_z$ may be created by an unpolarized pump. The effect also does not contradict time reversal symmetry: The spin polarization is formed here as a result of dissipative processes (energy and spin relaxation), which break time reversal symmetry, and would vanish at equilibrium [24,25]. Therefore, the kinetic spin-polarization of excitons discussed here is fundamentally different from the optical effects related to spatial dispersion, for which the photon wave vector plays a crucial role.

## Conclusion

We report on the emergence of significant degrees of circular polarization in the photoluminescence emitted from Moiré excitons in a MoSe$_2$/WSe$_2$ van der Waals heterostructure. The DCP reaches experimental values in excess of 0.4 under conditions of retained time reversal symmetry. This observation is qualitatively described by a kinetic model accounting for the relaxation of excitons from an intermediate state with rotated eigen-frames. The IX polarization emission presents a sensitive reaction on external magnetic fields, which can either enhance or quench the polarization, strongly suggests a kinetic origin arising from the dynamic relaxation of excitons from intermediate states with reduced symmetry and rotated eigen-frames. Our work displays a clear strategy to engineer chiral materials utilizing a twistronics-based approach, paving a way towards atomically thin emitters of circularly polarized radiation.

## Acknowledgements

We thank M. Florian, D. Erben, S. Kersch, and A.V. Poshakinskiy for useful discussion. We acknowledge funding via the DFG priority program via the projects SCHN1376 14-1, RE2974/26-1, 2244, and 2247 QM$^3$ (graduate school program). F.L. is grateful for his support from the CCDF of the University of Bremen. S.A.T. acknowledges support by Russian Science Foundation (grant 19-12-00051). T.V.S. acknowledges support by Russian Science Foundation (grant 19-12-00273; data analysis). ST acknowledges DOE-SC0020653, DMR 2111812, DMR 1933214, DMR 1904716 and ECCS 2052527 for material development and integration. K.W. and T.T. acknowledge support from the MEXT Element Strategy Initiative to Form Core Research Center, Grant Number JPMXP0112101001 and JSPS KAKENHI Grant Number JP20H00354.

# Supplementary Material: Intrinsic circularly-polarized exciton emission in a twisted van-der-Waals heterostructure

J. Michl[1], S.A. Tarasenko[2], F. Lohof[3], G. Gies[3], M. von Helversen[4], R. Sailus[5], S. Tongay[5], T. Taniguchi[6], K. Watanabe[7], T. Heindel[4], S. Reitzenstein[4], T. Shubina[2], S. Höfling[1], C. Anton-Solanas[1,8], C. Schneider[8]

[1]*Technische Physik, Physikalische Institut and Wilhelm Conrad Röntgen-Center for Complex Material Systems, Universität Würzburg, Am Hubland, D-97074 Würzburg, Germany*
[2]*Ioffe Institute, 194021 St. Petersburg, Russia*
[3]*Institute for Theoretical Physics, University of Bremen, D-28334 Bremen, Germany*
[4]*Institute of Solid-State Physics, Technische Universität Berlin, D-10623 Berlin, Germany.*
[5]*School for Engineering of Matter, Transport, and Energy, Arizona State University, Tempe, AZ, USA.*
[6]*International Center for Materials Nanoarchitectonics, National Institute for Materials Science, 1-1 Namiki, Tsukuba 305-0044, Japan*
[7]*Research Center for Functional Materials, National Institute for Materials Science, 1-1 Namiki, Tsukuba 305-0044, Japan*
[8]*Institute of Physics, Carl von Ossietzky University, 26129 Oldenburg, Germany.*

Earlier works have attributed the emergence of a multi-peak structure in the PL signal of slightly rotated van-der-Waals heterostructures to the emergence of Moiré trapped excitons [1]. To support this interpretation in our sample, we model the energy spectrum of IX excitons in the presence of a Moiré potential using a low-energy continuum model [2,3]. In this framework, an effective exciton Hamiltonian is formulated that describes the center-of-mass motion of the interlayer excitons in the twist-angle dependent Moiré potential landscape

$$H = -\frac{\hbar^2}{2M}\Delta_R + V^M(R). \quad [S1]$$

The first term describes the center-of-mass kinetic energy of the IX with $M = m_e + m_h$ its total mass, for which we use $m_e = 0.49 m_0$, $m_h = 0.35 m_0$, and $m_0$ the free electron mass. $V^M(R)$ is the Moiré modulation potential that the IXs experience due to the periodic variation of the local bandgap that results from change in stacking configuration of the bilayers along the Moire lattice. The Moiré potential is well approximated by an expansion over reciprocal Moiré lattice vectors $G_j$ from the first shell around the zeroth Brillouin zone,

$$V^M(R) \approx \sum_{j=1}^{3} 2V\cos(G_j \cdot R + \phi), \quad [S2]$$

with $V = 11.8$ meV, $\phi = 79.5°$, and the three $G_j$ span an angle of $120°$ among each other [2]. Here, we neglect the electrons and holes relative motion, noticing that the Moiré periodicity is sufficiently larger than the IX Bohr radius, and the IX is thus described as traversing the Moiré potential landscape as a composite particle.

In Fig 2(b) of the main text, we plot the IX Moiré exciton band structure as a function of momentum in the Moiré mini-Brillouin zone. At a rotation angle of 5°, Eq. (S1) yields a set of optical resonances at *K/K-* [4] that are separated by 20-40 meV. Due to the indirect nature of IX in twisted Moiré structures, it is important to note that these resonances are typically not observed in absorption, and the interpretation of PL spectra is intrinsically more difficult due to the influence of population effects [5]. The resonances inherit optical selection rules from the $C_3$-rotational symmetry of the Moiré IX potential, which results in left and $\sigma^-$ and $\sigma^+$ polarization for the lowest two resonances, respectively ( $\sigma^+$ and $\sigma^-$ for the time-reversal partners). For the identification of the resonances in the emission spectrum, we consider the selection rules on the basis of Dipole matrix elements for $\sigma^\pm$ and $z$ polarization, according to which the lower two resonances are bright for normal incidence, and we associate these with the two lower-energy peaks in the measures PL spectrum in the right panel of Fig. 2(b) of the main text. The third moiré-exciton band obeys selection rules that corresponds to linear polarized emission and a dipole normal to the bilayer plane. It is intrinsically dark for normal incidence, but contributes z-polarized emission as we can infer from evaluating the optical response for parallel incidence. Acknowledging that the theoretical model represents a perfectly homogeneous and ideal Moiré structure, and the fact that the wide aperture of the experimental setup collects some light away from normal incidence, we attribute the emission from the third peak at about 1.41 eV to contributions from the third Moiré exciton band.

**Supp. References**